\documentclass[11pt,simpleeqn,twoside,reqno]{amspaper}
\usepackage{amsmath}
\usepackage{amsfonts}
\usepackage{natbib}
\usepackage{times}
\bibpunct{(}{)}{,}{a}{}{,}
\usepackage[notheorems]{mathdef}
\newcounter{commentcounter}
{\end{list}}
\begin{document}
\thispagestyle{empty}
\noindent
Comment on:\\
{\bfseries Evaluating causal relations in neural systems: Granger
causality, directed transfer function and statistical assessment of
significance}\\
by Kaminski et al.
\bigskip

\noindent
Michael Eichler\footnote[1]{E-mail address: m.eichler@maastrichtuniversity.nl}\\
{\slshape\footnotesize
Department of Quantitative Economics, Maastricht University, P.O.~Box 616, NL-6200 MD Maastricht}
\bigskip

\noindent
The directed transfer function (DTF) introduced by \citet{kaminski91} is a well-known frequency-domain based measure for the interrelationships in multivariate time series. In the paper by \citet{kaminski01}, the authors claim
a relationship between the DTF and the concept of Granger causality. Here, Granger causality from one channel $X_i$ to another channel $X_j$ is defined in terms of a bivariate VAR model
\begin{align*}
X_i(t)&=\lsum_{u=1}^p A_{ii}(u)X_i(t-u)+\lsum_{u=1}^p
A_{ij}(u)X_j(t-u)+e_i(t)\\
X_j(t)&=\lsum_{u=1}^p A_{ji}(u)X_i(t-u)+\lsum_{u=1}^p
A_{jj}(u)X_j(t-u)+e_j(t),
\end{align*}
and $X_i$ is said to Granger cause $X_j$ if $A_{ji}(u)$ is nonzero
for some $u=1,\ldots,p$. We note that this bivariate notion of Granger
causality has been widely used \citep[e.g.,][]{fm85,goebel03,hesse03},
but for multivariate systems a more general notion of Granger causality
in terms of multivariate VAR models exists \citep[e.g.,][]{sims80a,
hsiao82,todaphilipps93,hayo99,eichlerpathdiagr,eichlerbrain}, which
is more in line with the original definition by
\citet{granger69,granger80,granger88}. As illustrated in \citet{eichlerdtf}, the DTF is neither a measure for this multivariate Granger causality nor actually for the bivariate notion described above. In the following, we show in detail the problems with the proof of \citet{kaminski01}.

For the proof of a relation between bivariate Granger causality and DTF,
the authors derive the bivariate autoregressive representation of two
components of a multivariate VAR(p) process (cf eqs (12) to (14)).
We note that the autoregressive representation of a weakly stationary
process is defined in terms of linear projections, which implies that
the error process $e(t)=(e_i(t),e_j(t))'$ is white noise, that is, the
errors at different time points are uncorrelated. In the frequency domain,
this implies that the spectral matrix of the error process is constant and equal
to $\mathbf\Sigma/2\pi$, where $\mathbf\Sigma=\var(e(t))$.

In the paper, the authors derive the bivariate autoregressive representation
(setting $i=1$ and $j=2$) expressed in the frequency domain
\begin{equation}
\label{biVAR}
\big[\mathbf{A}_{11}(\lam)-\mathbf{A}_{12}(\lam)
\mathbf{A}_{22}(\lam)^{-1}\mathbf{A}_{21}(\lam)\big]
\begin{pmatrix}
X_1(\lam)\\
X_2(\lam)
\end{pmatrix}=\begin{pmatrix}
E'_1(\lam)\\E'_2(\lam)
\end{pmatrix}
\end{equation}
(cf eqn (14)) with error process
\[
\begin{pmatrix}
E'_1(\lam)\\E'_2(\lam)
\end{pmatrix}
=\begin{pmatrix}
E_1(\lam)\\E_2(\lam)
\end{pmatrix}
-\mathbf{A}_{12}(\lam)\mathbf{A}_{22}(\lam)^{-1}
\begin{pmatrix} E_3(\lam)\\\vdots\\E_p(\lam)\end{pmatrix}.
\]
The spectral matrix $f_{e'_1e'_2}(\lam)$ of the error $e'(t)$ process
is given by
\[
\begin{split}
2\pi\,\mathbf{f}_{e'_1e'_2}(\lam)
=&\begin{pmatrix}
\Sigma_{11}&\Sigma_{12}\\
\Sigma_{21}&\Sigma_{22}
\end{pmatrix}
-\mathbf{A}_{12}(\lam)\mathbf{A}_{22}(\lam)^{-1}
\begin{pmatrix}
\Sigma_{31}&\Sigma_{32}\\\vdots&\vdots\\
\Sigma_{p1}&\Sigma_{p2}
\end{pmatrix}\\
&-\begin{pmatrix}
\Sigma_{13}&\ldots&\Sigma_{1p}\\
\Sigma_{23}&\ldots&\Sigma_{2p}
\end{pmatrix}
(\mathbf{A}_{22}(\lam)')^{-1}\mathbf{A}_{12}(\lam)'\\
&+\mathbf{A}_{12}(\lam)\mathbf{A}_{22}(\lam)^{-1}
\begin{pmatrix}
\Sigma_{33}&\ldots&\Sigma_{3p}\\\vdots&\ddots&\vdots\\
\Sigma_{p3}&\ldots&\Sigma_{pp}
\end{pmatrix}
(\mathbf{A}_{22}(\lam)')^{-1}\mathbf{A}_{12}(\lam)'
\end{split}
\]
Due to the frequency dependency of $\mathbf{A}_{11}(\lam)$,
$\mathbf{A}_{12}(\lam)$, and $\mathbf{A}_{22}(\lam)$,
this expression in general will not be constant over frequency and, thus,
cannot be the spectral matrix of a white noise process. Consequently,
the process $e'(t)=(e_1(t),e_2(t))'$ defined by
$E'(\lam)=(E'_1(\lam),E'_2(\lam))$ in general is not a white noise process
and \eqref{biVAR} is not the desired bivariate autoregressive representation.

That $e'(t)=(e_1(t),e_2(t))'$ indeed is not generally a white noise process
can be shown by a simple example. Consider a simple trivariate VAR(1) model
\[
\begin{split}
X_1(t)&=\alpha\,X_3(t-2)+\veps_1(t),\\
X_2(t)&=\beta\,X_3(t-1)+\veps_2(t),\\
X_3(t)&=\veps_3(t),
\end{split}
\]
where $\veps(t)=(\veps_1(t),\veps_2(t),\veps_3(t))$ is a white noise
process with mean zero and variance equal to the identity matrix.
On the one hand, we have
\[
\mathbf{A}(\lam)
=\begin{pmatrix}
1 & 0 & -\alpha\\
0 & 1 & -\beta\\
0 & 0 & 1
\end{pmatrix},
\]
and simple manipulations show that
\[
\mathbf{H}(\lam)=\mathbf{A}(\lam)^{-1}=\begin{pmatrix}
1 & 0 & \alpha\\
0 & 1 & \beta\\
0 & 0 & 1
\end{pmatrix},
\]
which implies that the DTF from channel $2$ to channel $1$ is zero.

On the other hand, the bivariate autoregressive representation is
given by the best predictor of $\tilde X(t)=(X_1(t),X_2(t))$
based on $\tilde X(t-1),\tilde X(t-2),\ldots$. It can be shown that
it is given by
\[
\begin{split}
X_1(t)&=
\SSS{\frac{\alpha\,\beta}{1+\beta^2}}\,X_2(t-1)+\tilde\veps_1(t),\\
X_2(t)&=\tilde\veps_2(t),\\
\end{split}
\]
where $\tilde\veps_2(t)=\veps_2(t)+\beta\,\veps_3(t-1)$ and
\[
\tilde\veps_1(t)=\veps_1(t)-\SSS{\frac{\alpha\beta}{1+\beta^2}}\,\veps_2(t-1)
+\SSS{\frac{\alpha}{1+\beta^2}}\,\veps_3(t-2).
\]
Note that $\tilde\veps(t)=(\tilde\veps_1(t),\tilde\veps_2(t))$ is indeed
a white noise process satisfying
\[
\mean\big(\tilde\veps(t)\tilde\veps(s)'\big)=0
\]
for all $t\neq s$. In particular, we have
\[
\cov(\tilde\veps_1(t-1),\tilde\veps_2(t))
=-\SSS{\frac{\alpha\beta}{1+\beta^2}}+\SSS{\frac{\alpha\beta}{1+\beta^2}}=0.
\]
It follows that $X_2$ bivariately Granger causes $X_1$ despite the fact
that the DTF is zero. Thus the example contradicts the
result by \citeauthor{kaminski01}.

We note that the error process $\tilde\veps$ in the above bivariate
representation differs from the error process $\veps'$ proposed by
\citeauthor{kaminski01}, which is of the form (written in the time domain)
\[
\begin{split}
\veps'_1(t)&=\veps_1(t)+\alpha\veps_3(t-1)\\
\veps'_2(t)&=\veps_2(t)+\beta\veps_3(t-2)
\end{split}.
\]
Obviously we have
\[
\mean(\veps'_1(t-1)\veps'_2(t))=\alpha\beta\neq 0,
\]
that is, the process $\veps'(t)=(\veps'_1(t),\veps'_2(t))$ is not
a white noise process as required by the autoregressive representation
used in the definition of Granger-causality. As a consequence, the
temporal dependence structure that is still hidden in the dependencies
of $\veps'$ is neglected when computing Granger-causality based on
the bivariate representation \eqref{biVAR}.

\bibliographystyle{stat}
\bibliography{papers,application}
\end{document}